# Geometry-Driven Magnetoelectric Coupling in Two-Dimensional Compensated Ferrimagnets

Peibo Xu,[1,3,+] Yixuan Che,[1,+] Haifeng Lv,[2,*] Xiaojun Wu,[2,3,*] and Jinlong Yang[1,2,3]

[1] *School of Emerging Technology and Hefei National Research Center for Physical Sciences at the Microscale, University of Science and Technology of China, Hefei, Anhui 230026, China*

[2]*State Key Laboratory of Precision and Intelligent Chemistry, School of Chemistry and Materials Science, and Collaborative Innovation Center of Chemistry for Energy Materials (iChEM), University of Science and Technology of China, Hefei, Anhui 230026, China*

[3]*Hefei National Laboratory, University of Science and Technology of China, Hefei, Anhui 230088, China*

[+] P.X. and Y.C. contributed equally to this work.

[*]Contact authors: xjwu@ustc.edu.cn (X.W.); hflv@ustc.edu.cn (H.L.)

The magnetoelectric coupling in compensated magnets enables stray-field-free manipulation of spin-splitting, holding great promise for spintronics, but inherently hindered by the symmetry mismatch between spatial-inversion-broken ferroelectricity and time-reversal-broken spin states. Here, based on a symmetry-decoupled analysis of magnetoelectric coupling in compensated magnets, we establish a geometry-driven spin-ferroelectric coupling mechanism in bilayer breathing kagome lattices. Within this geometric framework interlocking the out-of-plane electric polarization with cooperative intralayer structural distortions, we demonstrate that polarization switching drives a deterministic reversal of the global spin splitting. First-principles calculations on a prototype bilayer $Nb_3Cl_8$ successfully validate this mechanism, demonstrating the switching of spin-splitting states through an energetically feasible, asynchronous layer-by-layer transition pathway. Our proposed coupling originates from lattice geometry and structural symmetry, establishing a unique route toward switchable spin splitting in compensated ferrimagnets.

*Introduction*—Efficient electrical manipulation of spin-dependent electronic states is central to next-generation ultralow-power spintronics [1–3]. Among the various spin functionalities, spin splitting is particularly attractive owing to its direct impact on spin-dependent electronic and transport properties [4–6]. Fundamentally, however, the distinct transformation properties of electric polarization and spins under spatial-inversion ($P$) and time-reversal ($T$) symmetries forbid their direct coupling in the absence of mediating degrees of freedom [7]. To bridge this gap, magnetoelectric coupling provides a natural route for linking electric and spin degrees of freedom, offering a promising framework for switchable spin functionalities [6,8].

In conventional multiferroics, electric polarization and magnetic order can be coupled through mechanisms such as spin–orbit-driven inverse Dzyaloshinskii–Moriya interaction, spin–lattice-mediated exchange striction, and spin-dependent charge hybridization, thereby enabling electrical manipulation of magnetic states [6,9–14]. However, functionalities based on magnetic order manipulation or macroscopic magnetization typically generate detrimental stray fields that hinder device miniaturization and high-density integration [15,16]. In this context, compensated magnets have recently emerged as a viable alternative by combining vanishing net magnetization with robust spin polarization.

Recent advances in compensated magnetism have uncovered a broad family of altermagnets and fully compensated ferrimagnets [17–23]. In altermagnets, opposite-spin sublattices are related by rotational or mirror symmetries rather than inversion or translation operations, giving rise to momentum-dependent spin-split electronic structures intimately tied to the lattice symmetry [17–19]. Consequently, existing demonstrations of switchable spin splitting have been largely confined to altermagnetic systems, where ionic displacements, sliding ferroelectricity, or related structural transformations reorient the momentum-dependent alternating spin polarization [24–31].

Fully compensated ferrimagnets pose a fundamentally different challenge [21]. Unlike altermagnets, the spin-splitting in compensated ferrimagnets cannot be reversed through a simple symmetry reorientation of alternating spin polarization, and opposite-spin sublattices in compensated ferrimagnets are not related by any spatial symmetry operation. Consequently, reversing the spin-splitting state requires a collective interchange of the globally defined spin splitting across the Brillouin zone while preserving exact magnetic compensation. Magnetoelectric mechanisms capable of achieving such nonvolatile switching remain largely unexplored.

Here, we propose a geometry-driven spin–ferroelectric coupling mechanism in two-dimensional (2D) compensated ferrimagnets based on bilayer breathing kagome lattices. Through a symmetry-decoupled analysis, we demonstrate that this geometric mechanism manifests on two distinct levels: intralayer geometric breathing distortions switch the ferroelectric polarization, while interlayer geometric stacking configurations dictate magnetic compensation and the emergence of spin splitting. The synergy between these two geometric dimensions interlock ferroelectric polarization and spin-splitting states, enabling polarization reversal to drive deterministic switching of global spin splitting. Using bilayer $Nb_3Cl_8$ as a prototype realization, first-principles calculations confirm the proposed mechanism and reveal an energetically accessible asynchronous layer-by-layer switching pathway connecting opposite spin-splitting states.

*Symmetry design rules*—FIG. 1(a) illustrates a symmetry-decoupled magnetoelectric coupling framework for 2D

compensated magnets featuring layer-resolved spin orientation and electric polarization distribution, where the local spin-up and spin-down configurations are denoted by red and blue arrows, respectively, and the polarization state is mapped by its purple and orange coloring. To characterize the magnetoelectric properties of the system, we denote the net out-of-plane electric polarization by $P$ and the spin-splitting state by $S(k)$. For a given polarization state $P$, the momentum-resolved spin-splitting state $S(k)$ is defined as the energy difference between the corresponding spin-resolved bands,

$$(P, S(k)) = (P, E_{\uparrow}(k) - E_{\downarrow}(k)). \tag{1}$$

Consequently, the core objective of magnetoelectric coupling in such compensated magnets is to realize deterministic switching between the initial state ($P$, $S$) and its fully reversed state ($\bar{P}$, $\bar{S}$), inherently dictating

$$S(k) \xleftrightarrow{P\leftrightarrow\bar{P}} \bar{S}(k). \tag{2}$$

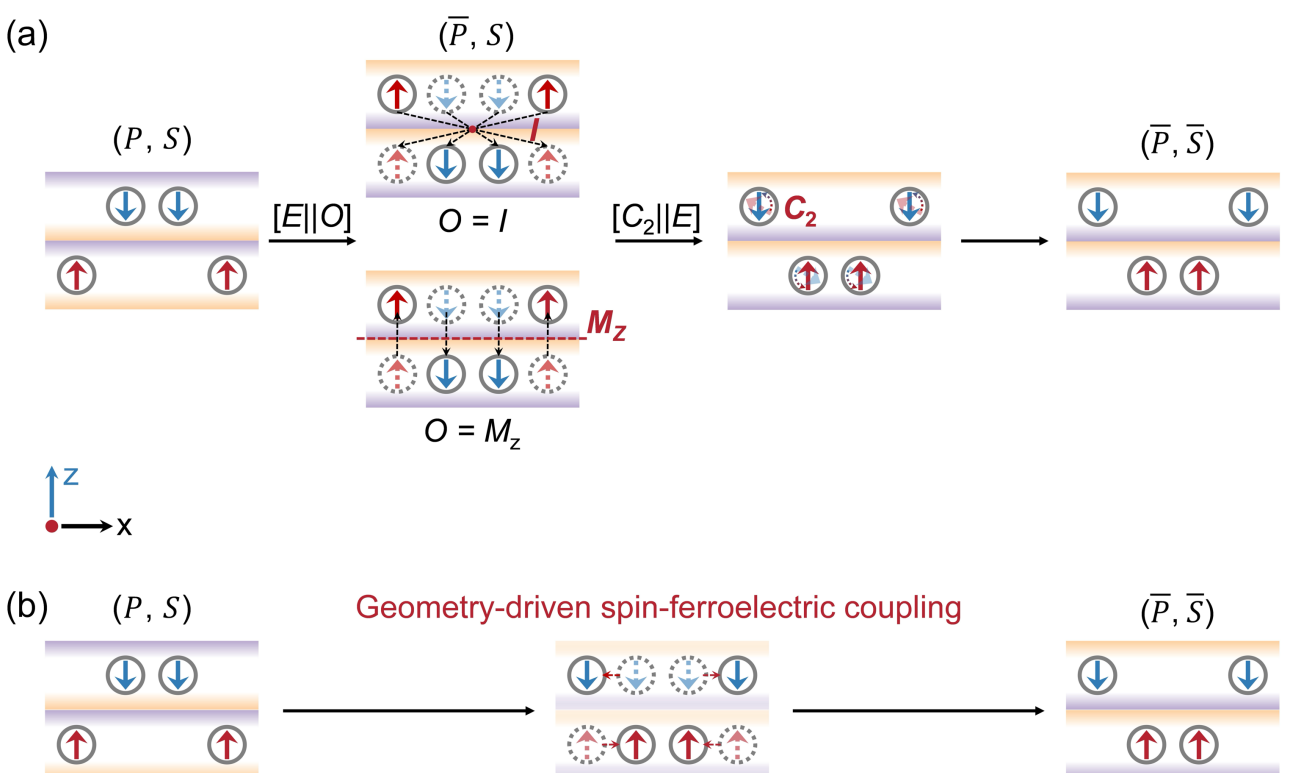

FIG. 1. Schematic illustration of magnetoelectric switching pathways in 2D compensated magnets with layer-resolved spin orientation and electric polarization distribution. The local spin-up and spin-down atoms are denoted by red and blue arrows, respectively, and the polarization state is mapped by its purple and orange coloring. (a) The symmetry-allowed theoretical pathways connecting the initial ($P$, $S$) state to the final ($\bar{P}$, $\bar{S}$) state. (b) The proposed geometry-driven spin-ferroelectric coupling mechanism. Synchronous in-plane atomic displacements and out-of-plane polarization flips within each layer drive a cooperative transition to the ($\bar{P}$, $\bar{S}$) state.

To resolve the symmetry origin of this coupled reversal, we adopt the spin-group representation $[R_s||R]$, in which spin-space ($R_s$) and real-space operations ($R$) are effectively decoupled in the absence of spin–orbit coupling (SOC) [32–34]. Within this framework, accessing the opposite state ($\bar{P}$, $\bar{S}$) dictates the reversal of both electric and spin polarization. In the real-space sector, an independent point-group operation $A$ ($A \in \{I, M_z, C_{2x(y)}$, or $S_{4z}\}$) is required to reverse the out-of-plane polarization:

$$[E||A](P, S(k)) = (\bar{P}, S(Ak)). \tag{3}$$

Among these operations, the subset $O \in \{I, M_z\}$ reverses $P$ while preserving the in-plane momentum $k$ (see details in Note S2), thereby driving the system from the initial ($P$, $S$) state into the intermediate ($\bar{P}, S$) state

$$(P, S) \xrightarrow{[E||O]} (\bar{P}, S). \tag{4}$$

To ultimately access the target counterpart ($\bar{P}$, $\bar{S}$) state, a pure spin-space twofold rotation $[C_2||E]$ is required to reverse the spin state with the identical $P$:

$$[C_2||E]E_{\uparrow}(k) = E_{\downarrow}(k), \tag{5}$$

Consequently, the complete transition pathway bridging the two opposite states is governed by the combined transformation:

$$(P, S) \xrightarrow{[C_2||O]=[E||O]\times[C_2||E]} (\bar{P}, \bar{S}). \tag{6}$$

Although symmetry permits the transformation $[C_2||O]$, its direct realization in real materials would generally require collective spin reversal and cross-layer ionic migration, rendering such a pathway impractical in real materials. Remarkably, comparison of the initial and final states reveals an alternative route: if each monolayer undergoes an in-plane atomic displacement synchronized with its out-of-plane polarization flip, the cooperative bilayer transition becomes symmetry-equivalent to the $[C_2||O]$ operation [FIG. 1(b)]. This unveils a geometry-driven spin-ferroelectric coupling mechanism, enabling deterministic switching of spin splitting in 2D compensated magnets.

The symmetry analysis above highlights a structural principle for geometry-driven spin-ferroelectric coupling: polarization reversal should be intrinsically coupled to a rearrangement of the spin-carrying units. Breathing kagome lattices provide a natural realization of this requirement. As an inversion-breaking trimerization distortion, the breathing mode in kagome lattices generates alternating expanded and contracted triangles, thereby breaking spatial inversion ($I$) symmetry and inducing spontaneous out-of-plane polarization ($P$) [FIG. 2(a)-(b)] [35–41]. Furthermore, the localized magnetic moments on the contracted trimers ensure that the polar-phase-driven rearrangement of the trimer network simultaneously reconfigures the spin-carrying units, thereby establishing an essential structural prerequisite for a robust intrinsic coupling independent of SOC.

While the breathing distortion naturally couples polarization reversal to the rearrangement of spin-carrying trimers, realizing the deterministic magnetoelectric coupling shown in FIG. 1(b) further requires interlayer symmetries that ensure magnetic compensation and robust spin splitting. In this context, 2D spin-layer-coupled magnets provide a versatile platform for realizing compensated magnetic states, where interlayer antiferromagnetic coupling naturally separates opposite spins into different atomic layers [42–47]. Accordingly, we construct bilayer breathing kagome models based on an intralayer ferromagnetic and interlayer antiferromagnetic configuration, as depicted in FIG. 1. To isolate the intrinsic magnetoelectric characteristics of discrete stacking geometries, we restrict our analysis to high-symmetry configurations with aligned $C_3$ rotational axes and neglect interlayer sliding.

We first categorize the bilayer systems based on the interlayer registry of the magnetic contracted trimers. As illustrated in FIG. 2(c) and 2(d), two distinct spin registries can be identified: (i) a staggered configuration, in which the magnetic trimers of adjacent layers are spatially shifted [FIG. 2(c)], and (ii) an aligned configuration, in which they are vertically superimposed [FIG. 2(d)]. For each registry, the interlayer polarization alignment further adopts either a ferroelectric (parallel) or antiferroelectric (antiparallel) states by flipping one monolayer. Cross-coupling the two polar configurations with the two alternative spin distributions yields four distinct stacking modes, denoted AB, A'B, AA, and A'A [FIG. 2(e)-(h)]. Here, the letter sequence (AB and AA) specifies the staggered (AB) or aligned (AA) spin distribution with compensated bilayer polarization, while the prime symbol (') represents a flipping operation that alters the monolayer's polarization direction.

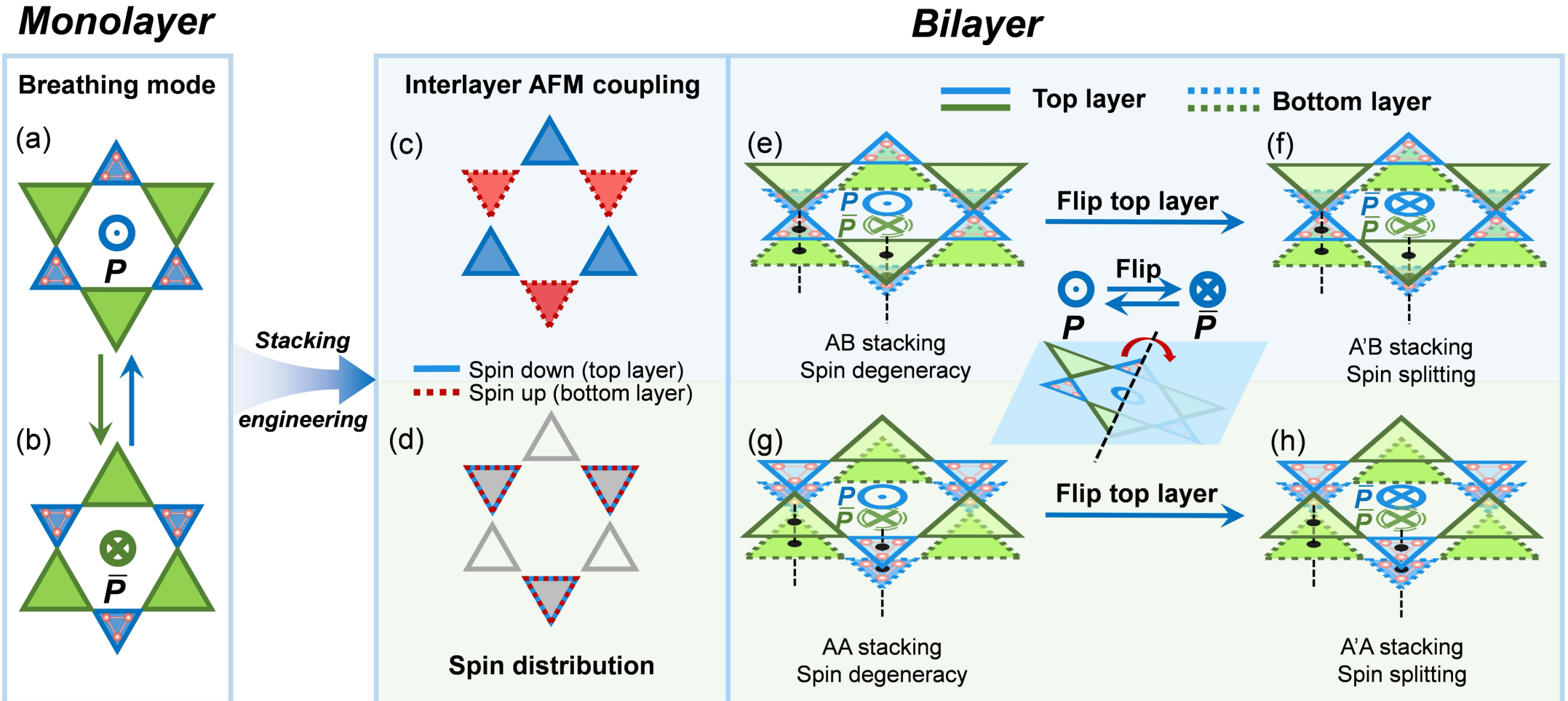

FIG. 2. Schematic of stacking engineering in breathing kagome lattices. (a, b) Switching out-of-plane polarization of monolayer breathing mode between (a) $P$ and (b) $\bar{P}$. Green and blue triangles denote expanded and contracted trimers formed by the magnetic ions (pink spheres), respectively, with magnetic moments localized on the contracted units. (c, d) Top views of two distinct spin distributions under intralayer FM and interlayer AFM coupling, characterized by spatially (c) staggered and (d) aligned spin arrangements between top and bottom layers. (e-h) Bilayer stacked breathing kagome lattices with distinct magnetic and polar states: The staggered series yields (e) an antiferromagnetic AB state and (f) a compensated ferrimagnetic A'B state upon flipping the top layer. Similarly, the aligned series results in (g) a Type-IV collinear AA state and (h) a compensated ferrimagnetic A'A state.

Collectively, these four stacking modes exhibit distinct magnetoelectric states determined by the symmetry operations connecting the interlayer sublattices. In the AB [FIG. 2(e)] and AA [FIG. 2(g)] configurations, the opposite-spin-sublattices are connected by $[C_2||I]$ or $[C_2||M_z]$ symmetries, respectively. These operations simultaneously enforce compensation of the out-of-plane electric dipoles and preserve spin degeneracy throughout the Brillouin zone. Notably, recent spin-group theory distinguishes these two operations in the classification of 2D collinear magnets: the $[C_2||I]$ operation characterizes conventional antiferromagnets, whereas $[C_2||M_z]$ defines the recently proposed type-IV class hosting time-reversal-symmetry-breaking responses under SOC [48,49].

In contrast, polarization reversal in the A'B [FIG. 2(f)] and A'A [FIG. 2(h)] configurations breaks these symmetry connections. The intrinsic polarization gives rise to a staggered electrostatic potential, imposing inequivalent on-site energies on the two magnetic monolayers and removing all spatial symmetries connecting opposite-spin sublattices. Consequently, the spin degeneracy is lifted across the entire Brillouin zone, generating a globally spin-split electronic structure while preserving exact magnetic compensation. These configurations therefore belong to the recently proposed class of fully compensated ferrimagnets combining robust spin splitting with a vanishing net magnetization [21].

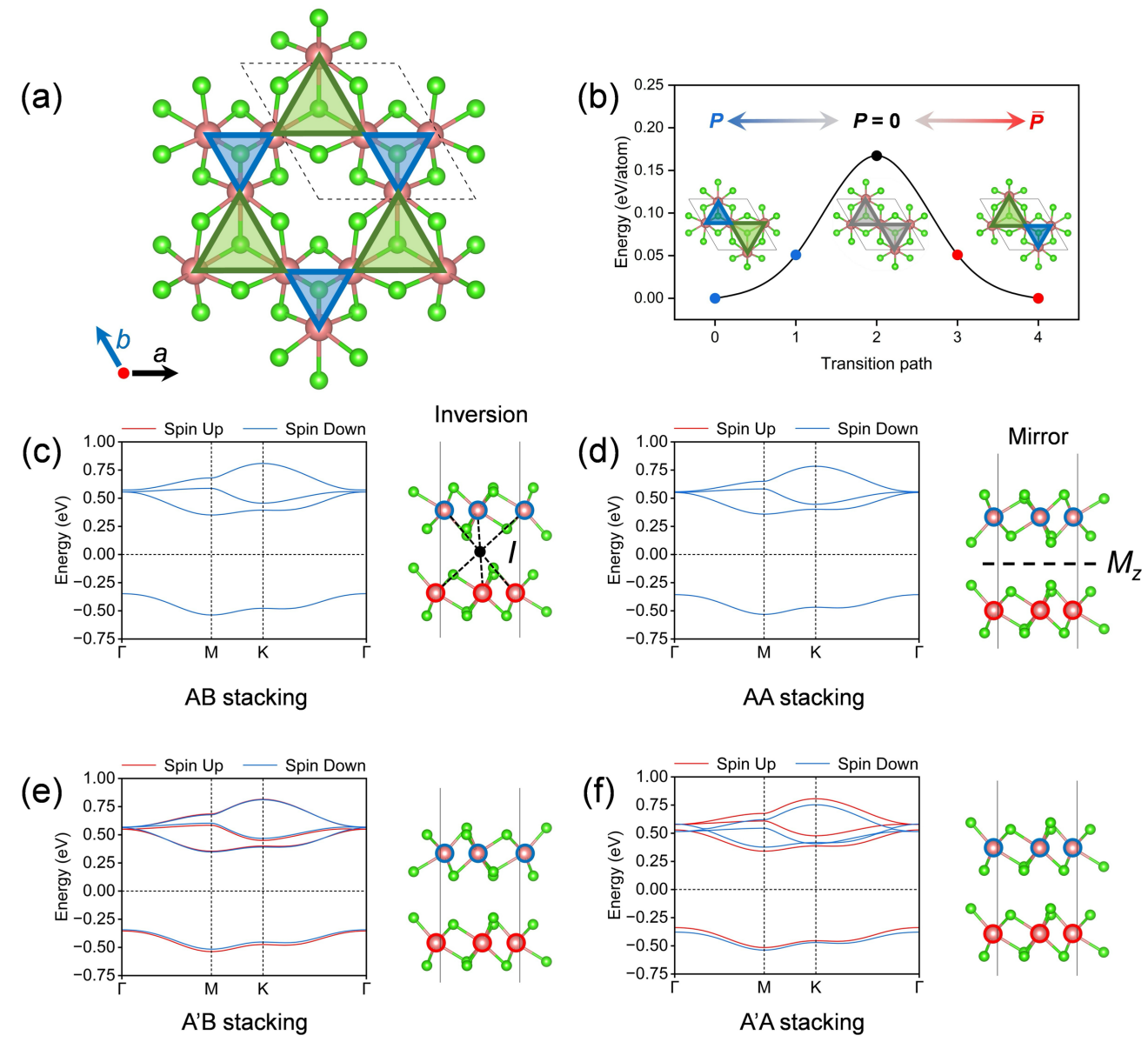

FIG. 3. Monolayer breathing mode switching and stacking-dependent spin electronic structure in $Nb_3Cl_8$ (a) Top view of the breathing kagome lattice of monolayer $Nb_3Cl_8$. Pink and green spheres represent Nb and Cl atoms, respectively. (b) The transition pathway for the breathing-mode switching in monolayer $Nb_3Cl_8$. (c-f) Stacking configurations and corresponding band structures of bilayer $Nb_3Cl_8$. The blue and red outlines denote spin-down in the top layer and spin-up in the bottom layer, respectively. The (c) AB and (d) AA

configurations preserve interlayer spatial inversion and mirror symmetries, respectively, enforcing strict spin degeneracy. In contrast, the polarization-flipped (e) A'B and (f) A'A configurations break these connecting symmetries, lifting the degeneracy to induce global spin splitting. Red and blue colors denote spin-up and spin-down channels, respectively.

*Materials Validation*—To validate the proposed geometry-driven spin–ferroelectric coupling mechanism, we identify bilayer $Nb_3Cl_8$ as a prototypical realization. As an experimentally confirmed material hosting the breathing kagome motif, bulk $Nb_3Cl_8$ features weak interlayer van der Waals coupling, enabling ready exfoliation into stable monolayers [39,50]. Previous studies have further predicted a ferromagnetic monolayer ground state [37,39,51], while the adjacent layers prefer antiferromagnetic interlayer coupling (see Supplemental Material [55], TABLE S1), naturally satisfying the layer-resolved spin configuration required by our model.

As shown in FIG. 3(a), monolayer $Nb_3Cl_8$ consists of trimerized Nb atoms with isolated $Nb_3Cl_{13}$ clusters composed of three edge-sharing octahedra, directly realizing the breathing kagome motif as illustrated in FIG. 2. The breathing distortion simultaneously generates spontaneous polarization and defines the spin-carrying trimers, thereby providing the structural interlocking required for geometry-driven spin–ferroelectric coupling. The ferroelectric switching pathway, calculated using the climbing-image nudged elastic band (CI-NEB) method, proceeds through a centrosymmetric non-polar saddle point with a calculated energy barrier of approximately 0.167 eV per atom, which is comparable with $Ta_3I_8$ (0.180 eV per atom) [52] and $W_3Br_8$ (0.191 eV per atom) [53] [FIG. 3(b), see methods in Supplemental Material [55], NOTE S1].

Upon extending the system to bilayers, the calculated electronic structures of bilayer $Nb_3Cl_8$ [FIG. 3(c)-(f)] faithfully reproduce the symmetry-dependent behaviors as illustrated in FIG. 2. The energetically favored AB configuration [TABLE S2, space group $P\overline{3}m1$] preserves interlayer spatial inversion and yields spin degenerate AFM state [FIG. 3(c) and S1], while the AA configuration [FIG. 3(d), space group $P\overline{6}m2$] constitutes a Type-IV collinear magnet protected by interlayer mirror operation [48]. In both cases, the electronic bands remain fully spin-degenerate in the absence of SOC (see Supplemental Material, FIG. S2 and S3), and the interlayer inversion (in AB configuration) and mirror (in AA configuration) symmetries enforce a zero net out-of-plane polarization (see Supplemental Material [55], TABLE S2).

By contrast, the polarization-flipped A'B [FIG. 3(e)] and A'A [FIG. 3(f)] configurations break these protective symmetries, leaving no spatial operation that relates the two opposite-spin sublattices. Consequently, both phases realize fully compensated ferrimagnetism: the symmetry breaking gives rise to an appreciable spin splitting throughout the entire Brillouin zone, while its sizable spin-channel gap of approximately 0.68 eV ensures the zero net magnetization through filling-enforced compensation. The maximum spin splitting near the Fermi level reaches 22 and 65 meV for the A'B and A'A stacking configurations, respectively. Notably, the magnetic easy axes for all configurations lie along the [110] direction, and the effect of SOC on the electronic band structures is negligible (see Supplemental Material, FIG. S2-S5 and TABLE S3).

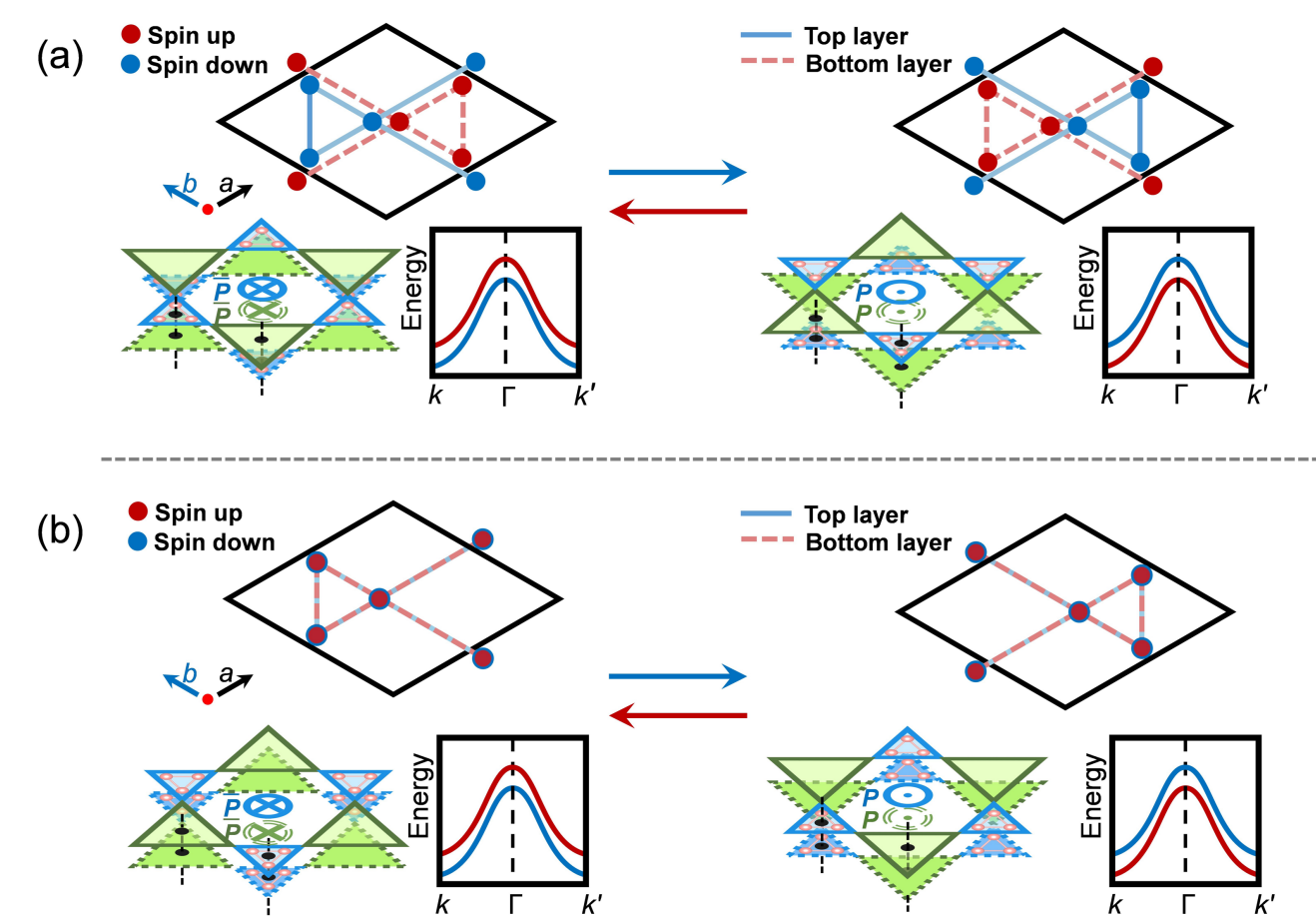

FIG. 4. Schematic of manipulation of spin splitting via bilayer breathing-mode switching. (a) Switching of A'B configuration with staggered spin arrangements. (b) Switching of A'A configuration with aligned spin arrangements.

Crucially, the A'A and A'B stackings satisfy the key symmetry conditions required for geometry-driven spin–ferroelectric coupling in bilayer breathing kagome lattices. At the intralayer level, the breathing distortion intrinsically links polarization reversal to the rearrangement of spin-carrying trimers. At the interlayer level, symmetry breaking together with antiferromagnetic exchange stabilizes a fully compensated magnetic state with robust global spin splitting. Consequently, the structural transition accompanying bilayer polarization reversal becomes symmetry-equivalent to a $[C_2||O]$ operation (FIG. 1), thereby enforcing deterministic switching between opposite spin-splitting states. FIG. 4 schematically illustrates the polarization switching states for the staggered A'B [FIG. 4(a)] and aligned A'A [FIG. 4(b)] configurations. Note that unlike sliding ferroelectricity, where polarization reversal is achieved through rigid interlayer translations [25,28,30,31], the present mechanism originates from an intralayer structural phase transition accompanying polarization reversal.

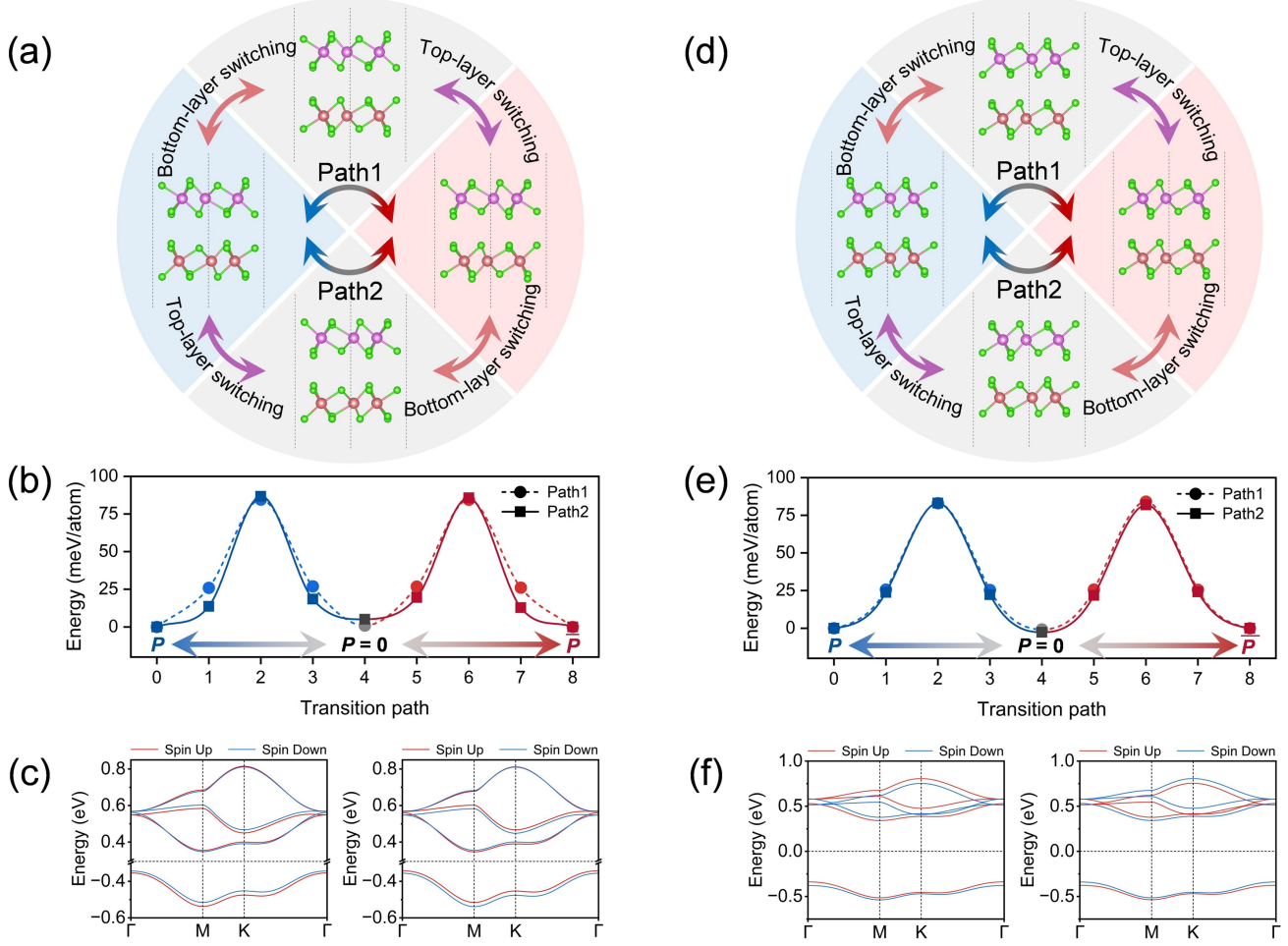
FIG. 5. Ferroelectric switching pathways and electronic structures of bilayer $Nb_3Cl_8$. (a–c) Results for the staggered A'B stacking configuration: (a) Schematic illustration of the asynchronous layer-by-layer switching pathways, where Path 1 and Path 2 denote the bottom-layer-first and top-layer-first switching sequences, respectively. The red and blue backgrounds denote opposite out-of-plane electric polarization states, while gray represents the nonpolar state. Purple and pink spheres represent the top-layer and bottom-layer Nb atoms, respectively, and green spheres represent Cl atoms. (b) Proposed transition pathways from the initial state to the reversed state obtained via CI-NEB calculations, passing through an intermediate nonpolar state ($P$=0). Blue and red colors denote opposite out-of-plane polarizations. (c) Spin-resolved electronic band structures for the initial state and the reversed state. The red and blue curves denote the spin-up and spin-down channels, respectively. (d–f) Corresponding results for the aligned A'A stacking configuration.

To examine the microscopic switching dynamics associated with the proposed geometry-driven spin–ferroelectric coupling, we performed CI-NEB calculations for polarization switching in bilayer $Nb_3Cl_8$. As illustrated in FIG. 5, both the staggered A'B [FIG. 5(a)-5(c)] and aligned A'A [FIG. 5(d)-5(f)] stackings favor an asynchronous layer-by-layer structural switching process. Such sequential switching reduces the structural distortion required at each stage and therefore lowers the associated transition barrier.

The sequence nature of the transition splits into two symmetry-related pathways, denoted as Path 1 and Path 2, corresponding to the bottom-layer-first and top-layer-first switching, respectively [FIG. 5(a) and 5(d)]. In both stacking configurations, the switching pathway proceeds through an intermediate nonpolar ($P$ = 0) configuration. The two pathways exhibit nearly identical energetics, with transition barriers of approximately 85 meV per atom for both stacking modes. This value is lower than that of $Sc_2CO_2$ (104 meV per atom) [54] and comparable with bilayer GeS (78 meV per atom) [55], suggesting the potential kinetic accessibility of this switching under experimentally relevant conditions. Importantly, the proposed geometry-driven spin-ferroelectric coupling mechanism is explicitly verified by the electronic band structures of the initial and opposite states [FIG. 5(c) and 5(f)]. For the A'B and A'A configurations, the structural transformations driven by the bilayer polarization reversal are symmetry-equivalent to applying $[C_2||M_z]$ and $[C_2||I]$ operations, respectively (see Supplemental Material, FIG. S6). Consequently, this symmetry mapping directly manifests as the reversal of spin splitting observed between the two states.

*Conclusion*—In summary, we establish a symmetry-decoupled framework for magnetoelectric coupling in 2D compensated magnets and propose a geometry-driven spin–ferroelectric coupling mechanism in fully compensated ferrimagnets based on bilayer breathing kagome lattices. Symmetry analysis reveals that the global structural transition accompanying the bilayer polarization reversal is symmetry-equivalent to a $[C_2||O]$ operation, which inherently enforces a simultaneous flip of the associated spin polarization. Using the bilayer $Nb_3Cl_8$ as a prototypical realization, first-principles calculations confirm that cooperative structural distortions enable a deterministic reversal of global spin-splitting states. By operating independently of relativistic spin-orbit interactions, it establishes a unique design principle for switchable spin splitting in compensated ferrimagnets and provides a promising route toward nonvolatile spintronic functionalities.

*Acknowledgements*—The research is supported by the National Natural Science Foundation of China (Grants Nos. 22225301, 22303092, and 22321001), the Quantum Science and Technology-National Science and Technology Major Project (Grant No. 2021ZD0303302), the Fundamental Research Funds for the Central Universities (Grants No. 20720250005, WK2490000001, and WK2490000002), the open research fund of Key Laboratory of Precision and Intelligent Chemistry, the robotic AI-Scientist platform of Chinese Academy of Sciences, and the Super Computer Center of USTCSCC and SCCAS.

*References*

[1] I. Žutić, J. Fabian, and S. Das Sarma, Spintronics: Fundamentals and applications, Rev. Mod. Phys. **76**, 323 (2004).
[2] S. Manipatruni, D. E. Nikonov, C.-C. Lin, T. A. Gosavi, H. Liu, B. Prasad, Y.-L. Huang, E. Bonturim, R. Ramesh, and I. A. Young, Scalable energy-efficient magnetoelectric spin–orbit logic, Nature **565**, 35 (2019).
[3] B. Dieny et al., Opportunities and challenges for spintronics in the microelectronics industry, Nat. Electron. **3**, 446 (2020).
[4] K. F. Mak, J. Shan, and D. C. Ralph, Probing and controlling magnetic states in 2D layered magnetic materials, Nat. Rev. Phys. **1**, 646 (2019).
[5] A. Fert, R. Ramesh, V. Garcia, F. Casanova, and M. Bibes, Electrical control of magnetism by electric field and current-induced torques, Rev. Mod. Phys. **96**, 015005 (2024).
[6] S. Dong, H. Xiang, and E. Dagotto, Magnetoelectricity in multiferroics: a theoretical perspective, Natl. Sci. Rev. **6**, 629 (2019).
[7] S. Dong, J. M. Liu, and S. W. Cheong, Multiferroic materials and magnetoelectric physics: Symmetry,

entanglement, excitation, and topology, Adv. Phys. **64**, 519 (2015).
[8] N. A. Spaldin and R. Ramesh, Advances in magnetoelectric multiferroics, Nat. Mater. **18**, 203 (2019).
[9] D. Khomskii, Classifying multiferroics: Mechanisms and effects, Physics **2**, 20 (2009).
[10] H. Katsura, N. Nagaosa, and A. V. Balatsky, Spin Current and Magnetoelectric Effect in Noncollinear Magnets, Phys. Rev. Lett. **95**, 057205 (2005).
[11] I. A. Sergienko and E. Dagotto, Role of the Dzyaloshinskii-Moriya interaction in multiferroic perovskites, Phys. Rev. B **73**, 094434 (2006).
[12] M. Fiebig, T. Lottermoser, D. Meier, and M. Trassin, The evolution of multiferroics, Nat. Rev. Mater. **1**, 16046 (2016).
[13] H. Yu, J. Ji, W. Luo, X. Gong, and H. Xiang, Recent Advances in Unconventional Ferroelectrics and Multiferroics, Adv. Mater. e07070 (2025).
[14] C. Tang and A. Du, Perspective on computational design of two-dimensional materials with robust multiferroic coupling, Appl. Phys. Lett. **122**, 130502 (2023).
[15] T. Jungwirth, X. Marti, P. Wadley, and J. Wunderlich, Antiferromagnetic spintronics, Nature Nanotech. **11**, 231 (2016).
[16] V. Baltz, A. Manchon, M. Tsoi, T. Moriyama, T. Ono, and Y. Tserkovnyak, Antiferromagnetic spintronics, Rev. Mod. Phys. **90**, 015005 (2018).
[17] L. Šmejkal, J. Sinova, and T. Jungwirth, Beyond Conventional Ferromagnetism and Antiferromagnetism: A Phase with Nonrelativistic Spin and Crystal Rotation Symmetry, Phys. Rev. X **12**, 031042 (2022).
[18] L. Šmejkal, J. Sinova, and T. Jungwirth, Emerging Research Landscape of Altermagnetism, Phys. Rev. X **12**, 040501 (2022).
[19] L. Bai, W. Feng, S. Liu, L. Šmejkal, Y. Mokrousov, and Y. Yao, Altermagnetism: Exploring New Frontiers in Magnetism and Spintronics, Adv. Funct. Mater. **34**, 2409327 (2024).
[20] Y. Che, H. Lv, X. Wu, and J. Yang, Realizing altermagnetism in two-dimensional metal–organic framework semiconductors with electric-field-controlled anisotropic spin current, Chem. Sci. **15**, 13853 (2024).
[21] Y. Liu, S.-D. Guo, Y. Li, and C.-C. Liu, Two-Dimensional Fully Compensated Ferrimagnetism, Phys. Rev. Lett. **134**, 116703 (2025).
[22] W. Zhao, X. Zhou, Z. Guo, T. Zhu, J. Chen, H. Li, Z. Cheng, X. Wang, and W. Wang, Multiferroic phase transition between multiple types of collinear compensated magnets, Nat Commun (2026).
[23] Y. Che, Y. Guo, H. Lv, X. Wu, and J. Yang, Symmetry-Driven Multiferroic Altermagnetism in Two-Dimensional Materials, J. Am. Chem. Soc. **148**, 5125 (2026).
[24] M. Gu, Y. Liu, H. Zhu, K. Yananose, X. Chen, Y. Hu, A. Stroppa, and Q. Liu, Ferroelectric Switchable Altermagnetism, Phys. Rev. Lett. **134**, 106802 (2025).
[25] W. Sun, C. Yang, W. Wang, Y. Liu, X. Wang, S. Huang, and Z. Cheng, Proposing Altermagnetic-Ferroelectric Type-III Multiferroics with Robust Magnetoelectric Coupling, Adv. Mater. **37**, 2502575 (2025).
[26] X. Duan, J. Zhang, Z. Zhu, Y. Liu, Z. Zhang, I. Žutić, and T. Zhou, Antiferroelectric Altermagnets: Antiferroelectricity Alters Magnets, Phys. Rev. Lett. **134**, 106801 (2025).
[27] Z. Zhu, X. Duan, J. Zhang, B. Hao, I. Žutić, and T. Zhou, Two-Dimensional Ferroelectric Altermagnets: From Model to Material Realization, Nano Lett. **25**, 9456 (2025).
[28] Y. Zhu, M. Gu, Y. Liu, X. Chen, Y. Li, S. Du, and Q. Liu, Sliding Ferroelectric Control of Unconventional Magnetism in Stacked Bilayers, Phys. Rev. Lett. **135**, 056801 (2025).
[29] S. Wang, W.-W. Wang, J. Fan, X. Zhou, X.-P. Li, and L. Wang, Two-Dimensional Dual-Switchable Ferroelectric Altermagnets: Altering Electrons and Magnons, Nano Lett. **25**, 14618 (2025).
[30] W. Sun, W. Wang, C. Yang, S. Huang, and Z. Cheng, A unified symmetry framework for spin–ferroelectric coupling in altermagnetic multiferroics, Nat Commun **17**, 3101 (2026).
[31] W. Sun, W. Wang, C. Yang, S. Huang, N. Ding, S. Dong, and Z. Cheng, Designing Spin Symmetry for Altermagnetism with Strong Magnetoelectric Coupling, Advanced Science **12**, e03235 (2025).
[32] D. B. Litvin, Spin point groups, Acta Cryst A **33**, 279 (1977).
[33] P. Liu, J. Li, J. Han, X. Wan, and Q. Liu, Spin-Group Symmetry in Magnetic Materials with Negligible Spin-Orbit Coupling, Phys. Rev. X **12**, 021016 (2022).
[34] X. Chen, J. Ren, Y. Zhu, Y. Yu, A. Zhang, P. Liu, J. Li, Y. Liu, C. Li, and Q. Liu, Enumeration and Representation Theory of Spin Space Groups, Phys. Rev. X **14**, 031038 (2024).
[35] S. Gao et al., Discovery of a Single-Band Mott Insulator in a van der Waals Flat-Band Compound, Phys. Rev. X **13**, 041049 (2023).
[36] Y. Wang, H. Wu, G. T. McCandless, J. Y. Chan, and M. N. Ali, Quantum states and intertwining phases in kagome materials, Nat. Rev. Phys. **5**, 635 (2023).
[37] Y. Xie, K. Ji, J. He, X. Shen, D. Wang, and J. Zhang, Manipulation of Topology by Electric Field in Breathing Kagome Lattice, Phys. Rev. Lett. **135**, 056701 (2025).
[38] R. Schaffer, Y. Huh, K. Hwang, and Y. B. Kim, Quantum spin liquid in a breathing kagome lattice, Phys. Rev. B **95**, 054410 (2017).
[39] Z. Sun et al., Observation of Topological Flat Bands in the Kagome Semiconductor $Nb_3Cl_8$, Nano Lett. **22**, 4596 (2022).
[40] S. Regmi et al., Spectroscopic evidence of flat bands in breathing kagome semiconductor Nb3I8, Commun. Mater. **3**, 100 (2022).
[41] H. Zhang et al., Topological Flat Bands in 2D Breathing-Kagome Lattice Nb3TeCl7, Adv. Mater. **35**, 2301790 (2023).
[42] Y. Liu, J. Yu, and C.-C. Liu, Twisted Magnetic Van der Waals Bilayers: An Ideal Platform for Altermagnetism, Phys. Rev. Lett. **133**, 206702 (2024).

[43] Y. Che, H. Lv, X. Wu, and J. Yang, Bilayer Metal–Organic Framework Altermagnets with Electrically Tunable Spin-Split Valleys, J. Am. Chem. Soc. **147**, 14806 (2025).
[44] H. Lv, Y. Niu, X. Wu, and J. Yang, Electric-Field Tunable Magnetism in van der Waals Bilayers with A-Type Antiferromagnetic Order: Unipolar versus Bipolar Magnetic Semiconductor, Nano Lett. **21**, 7050 (2021).
[45] S.-D. Guo, Y. Liu, J. Yu, and C.-C. Liu, Valley polarization in twisted altermagnetism, Phys. Rev. B **110**, L220402 (2024).
[46] M. Su, D. Zhang, H. Ye, G. P. Zhang, M. Gu, and J. Wang, Interlayer-sliding controlled magneto-optical effect and ferrovalley in a fully compensated ferrimagnetic bilayer, Phys. Rev. B **112**, 195427 (2025).
[47] R.-W. Zhang, C. Cui, R. Li, J. Duan, L. Li, Z.-M. Yu, and Y. Yao, Predictable Gate-Field Control of Spin in Altermagnets with Spin-Layer Coupling, Phys. Rev. Lett. **133**, 056401 (2024).
[48] L. Bai, R.-W. Zhang, W. Feng, and Y. Yao, Anomalous Hall Effect in Type IV 2D Collinear Magnets, Phys. Rev. Lett. **135**, 036702 (2025).
[49] M. Tian, C. Cui, Z. Zhang, J. Duan, W. Feng, and R.-W. Zhang, Symmetry Classification of Altermagnetism and Emergence of Type-IV Magnetism in Two Dimensions, Phys. Rev. Lett. **136**, 206701 (2026).
[50] C. M. Pasco, I. El Baggari, E. Bianco, L. F. Kourkoutis, and T. M. McQueen, Tunable Magnetic Transition to a Singlet Ground State in a 2D van der Waals Layered Trimerized Kagomé Magnet, ACS Nano **13**, 9457 (2019).
[51] J. Jiang, Q. Liang, R. Meng, Q. Yang, C. Tan, X. Sun, and X. Chen, Exploration of new ferromagnetic, semiconducting and biocompatible Nb3X8 (X = cl, br or I) monolayers with considerable visible and infrared light absorption, Nanoscale **9**, 2992 (2017).
[52] J. Lu, H. Chen, X. Zhao, G. Hu, X. Yuan, and J. Ren, Chiral breathing-valley locking in two-dimensional kagome lattice Ta3I8, Applied Physics Letters **124**, 072101 (2024).
[53] Y. Liu, F. Chen, G. Hu, X. Zhao, X. Yuan, and J. Ren, Tunable layer-locked valley Hall effect in ferroelectric-magnetic-valley coupled breathing Kagome bilayer W3Br8, Applied Physics Letters **126**, 133101 (2025).
[54] A. Chandrasekaran, A. Mishra, and A. K. Singh, Ferroelectricity, Antiferroelectricity, and Ultrathin 2D Electron/Hole Gas in Multifunctional Monolayer MXene, Nano Lett. **17**, 3290 (2017).
[55] B. Xu, J. Deng, X. Ding, J. Sun, and J. Z. Liu, Van der Waals force-induced intralayer ferroelectric-to-antiferroelectric transition via interlayer sliding in bilayer group-IV monochalcogenides, npj Comput. Mater. **8**, 47 (2022).
[56] See Supplemental Material for the details of computational methods, symmetry analysis and supplemental calculations on bilayer $Nb_3Cl_8$, which includes Ref. [56-68].
[57] R. G. Parr, Density Functional Theory, Annu. Rev. Phys. Chem. 34, 631 (1983).
[58] G. Kresse and J. Furthmüller, Efficient iterative schemes for ab initio total-energy calculations using a plane-wave basis set, Phys. Rev. B 54, 11169 (1996).
[59] G. Kresse and J. Furthmüller, Efficiency of ab-initio total energy calculations for metals and semiconductors using a plane-wave basis set, Comput. Mater. Sci. 6, 15 (1996).
[60] S. Grimme, J. Antony, S. Ehrlich, and H. Krieg, A consistent and accurate ab initio parametrization of density functional dispersion correction (DFT-D) for the 94 elements H-Pu, J. Chem. Phys. 132, 154104 (2010).
[61] S. Grimme, S. Ehrlich, and L. Goerigk, Effect of the damping function in dispersion corrected density functional theory, J. Comput. Chem. 32, 1456 (2011).
[62] J. P. Perdew, K. Burke, and M. Ernzerhof, Generalized Gradient Approximation Made Simple, Phys. Rev. Lett. 77, 3865 (1996).
[63] A. I. Liechtenstein, V. I. Anisimov, and J. Zaanen, Density-functional theory and strong interactions: Orbital ordering in Mott-Hubbard insulators, Phys. Rev. B 52, R5467 (1995).
[64] P. E. Blöchl, Projector augmented-wave method, Phys. Rev. B 50, 17953 (1994).
[65] G. Kresse and D. Joubert, From ultrasoft pseudopotentials to the projector augmented-wave method, Phys. Rev. B 59, 1758 (1999).
[66] H. J. Monkhorst and J. D. Pack, Special points for Brillouin-zone integrations, Phys. Rev. B 13, 5188 (1976).
[67] C. Zener, Classical Theory of the Temperature Dependence of Magnetic Anisotropy Energy, Phys. Rev. 96, 1335 (1954).
[68] G. Henkelman, B. P. Uberuaga, and H. Jónsson, A climbing image nudged elastic band method for finding saddle points and minimum energy paths, J. Chem. Phys. 113, 9901 (2000).
[69] S. Smidstrup, A. Pedersen, K. Stokbro, and H. Jónsson, Improved initial guess for minimum energy path calculations, J. Chem. Phys. 140, 214106 (2014).